\documentclass[%
 reprint,
 amsmath,amssymb,
 aps,
]{revtex4-1}

\usepackage{graphicx}
\usepackage{dcolumn}
\usepackage{bm}
\usepackage{bm}

\begin{document}

\preprint{APS/123-QED}

\title{Modified Bohm's Dynamics: A Stable Version}

\author{M. J. Kazemi}
\email{kazemi.j.m@gmail.com}
\author{M. Mashhadi}%
\author{M. H. Barati}%
\affiliation{ Department of Physics, Faculty of Science, Shahid Beheshti University, Tehran 19839, Iran}
\affiliation{Department of Physics, Faculty of Science,Qom University, Qom, Iran}
\affiliation{Department of Physics, Faculty of Science, Kharazmi University, Tehran, Iran}

\date{\today}

\begin{abstract}
Recently, it has been shown that the quantum equilibrium distribution in the original Bohm's model is unstable and so it isn't a tenable physical theory [Proc. R. Soc. A $\bm{470}$ 20140288 (2014)]. In this paper we show that a natural modification of the Bohm's quantum force leads to a stable quantum equilibrium without any change to statistical predictions of this model in equilibrium state.  Moreover, it is shown that an initial non-equilibrium phase space distribution relaxes to equilibrium distribution via  a coarse grained H-theorem. But before end of the relaxation process, for example in the early universe, this modified quantum force can lead to the new predictions beyond standard quantum mechanics.
 
\begin{description}

\item[PACS numbers]
03.65.Ta,  05.10.Gg

\end{description}
\end{abstract}

\pacs{Valid PACS appear here}
\maketitle


    \emph{1. Introduction-.} Bohmian mechanics is a completely coherent casual theory for describing quantum systems which introduced by L. de Broglie in 1927 \cite{L. de Broglie} and reformulated and developed by D. Bohm in 1952 \cite{D. Bohm52-1}. In this theory, unlike the orthodox quantum mechanics, the quantum particles have an actual position in each moment and moving along  well defined continuous trajectories under effect of the $\psi$-field. In non-relativistic Bohmian mechanics, the time evolution of the wave function, for a system of N spineless particles, is given by Schr\"{o}dinger  equation: 
\begin{equation}
i\hbar \frac{\partial \psi {\rm (}{{x}},t{\rm )}}{\partial t}{\rm =-}\sum^N_{i=1}{\frac{{\hbar }^{{\rm 2}}}{{\rm 2}m_i}{\nabla }^2_i\psi }{\rm +}V{\rm (}{x}{\rm )}\psi, 
\end{equation}
Where $m_i$ is the mass of $i$-th particle and ${x}=(\bm{x}_1,...,\bm{x}_N )\in{\mathbb R}^{3N}$. But for describing particles' motion, there are two approaches \cite{Quantum Theory of Motion,J. S. Bell1987, D. Dürr1992}: de Broglie's first-order dynamics and Bohm's second-order dynamics. In the de Broglie's dynamics, the state of the system is completely determined by the wave function $\psi $ and particles' positions $ {{ X}}=(\bm X_1,...,\bm X_N)  $,  while  the law of particles motion is given by  "guiding equation":
\begin{equation}
m\frac{d{{X}}(t)}{dt}={\left.{\nabla }S(x,t)\right|}_{{x}{\mathbf =}{X}(t)}, 
\end{equation}
Where $ S$ is the phase of the wave function,  $\psi=Re^{iS/\hbar} $ and $m$ is defined as $m= $ diag$(m_1,...,m_N)$ and $\nabla=(\nabla_1,...,\nabla_N)$. But in the Bohm's reformulation of the de Broglie's model, the state of the system is determined by wave function, particles' position and also particles' velocity. In this model, the particles' law of motion is given by Newton's law with an additional "quantum potential":
\begin{equation}
m\frac{d^2{{ X}}(t)}{dt^2}={\left.-{\nabla }(V+Q)\right|}_{{x}{=}{X}(t)},
\end{equation}
Where  $Q$  is defined as  $Q=\sum^N_{i=1}{(-{\hbar}^{2}/2{m}_{i}){{\nabla}_{i}}^{2}R/R}$.

Both of the de Broglie's and Bohm's models are deterministic theories  and so the "probability concept" is not considered as an intrinsic property in these models: i.e. the stochastic characteristics of the quantum phenomena  must be explained by the aid of the stochastic initial conditions \cite{Quantum Theory of Motion, D. Dürr1992}. It is well known that the empirical predictions of the standard quantum mechanics follow from  the de Broglie's dynamics, if it is assumed that the initial positions of particles is distributed as  
\begin{equation}
\rho(x;t_0) ={\left|\psi \right|}^2.
\end{equation}
The distribution $\rho=|\psi|^2$ is called "quantum equilibrium distribution". It is an elementary consequence of the de Broglie's dynamics that the quantum equilibrium distribution  is preserved in time: if $\rho=|\psi|^2$ at $t_0$ then $\rho=|\psi|^2$ at all future times. This property is referred to as "equivariance" \cite{D. Dürr1992}. Similarly, if the initial distribution of particles in phase space be as 
\begin{equation}
f_B\left({x},{p};t_0\right)=R^2\delta ({p}-\nabla S),
\end{equation}
Where $p=(\bm p_1,...,\bm p_N)$, then the empirical predictions of the standard quantum mechanics follow from the Bohm's dynamics. We name $f_B(x,p)=R^2\delta ({p}-\nabla S)$ as "Bohm's quantum equilibrium distribution". The Bohm's quantum equilibrium  is  preserved in time in the Bohm's dynamics same as $\rho=|\psi|^2$ in  the de Broglie's dynamics (equivariance) \cite{Quantum Theory of Motion}. The equivariance of the Bohm's quantum equilibrium $f_B(x,p)$, ensures the validity of the Born rule and guiding equation in all times. So in the quantum equilibrium conditions, both of the models lead to the same trajectories and so are equivalent. 
But it should be noted that in the both formulations of the Bohmian mechanics, the wave function $\psi$ is considered as an actual field in the configuration space which affects on the particle's motion, so at the fundamental level, $\psi$ is not a priori related to probability notion, even through initial condition. Firstly, W. E. Pauli objected that in a fundamentally deterministic theory, taking a particular distribution $ \rho=|\psi|^2$ as an initial condition is not reasonable \cite{W. Pauli}. Notwithstanding this objection, D. Bohm had already suggested in his original papers that the distribution $ \rho=|\psi|^2$ was  similar to thermal equilibrium distribution in ordinary statistical mechanics, and must be derived by suitable statistical-mechanical arguments \cite{D. Bohm1953, D. Bohm1954}. In this regard, in 1991, A.Valentini showed in the de Broglie's dynamics, an arbitrary non-equilibrium distributions relax to quantum equilibrium distribution,  $\rho \to |\psi|^2$, based on a "sub-quantum H-theorem" \cite{A. Valentini1991}. Moreover, in recent years, the numerical simulations have shown that an initial non-equilibrium distribution $\rho \neq |\psi|^2$ rapidly approaches to $|\psi|^2$ on a coarse-grained level \cite{A. Valentini2005, S. Colin2010, A. Valentini2012, S. Colin2012}. But the concept of "non-equilibrium state"  in the Bohm's dynamics is wider than  the de Broglie dynamics; because in the former, the particles' velocities may be deviate from $\nabla S/m$, beside the fact that  $\rho$ may be deviate from $|\psi|^2$. The question that arises at this point is whether or not the  non-equilibrium distributions in Bohm's dynamics, $f\ne R^2\delta ({p}-\nabla S)$, tend to relax towards equilibrium distribution $f_B$. Because of two following facts, an initial distribution with non-zero Lebesgue measure support will not be able to relax to Bohm's quantum equilibrium \cite{S. Colin2014}: I) The support of the Bohm's quantum equilibrium, has zero Lebesgue measure in phase space, II) The phase-space volume is conserved in  the Bohm's dynamics. 
In addition, recently it has been shown that the quantum equilibrium in the Bohm's dynamics - unlike de Broglie dynamics -is unstable i.e.  a small perturbation  to the equilibrium rapidly grows with time \cite{S. Colin2014}. To emphasize this instability, S. Goldstein and W. Struyve, showed that for various wave functions, the particles tend to escape the wave packet in Bohm's dynamics \cite{Sheldon Goldstein2015}. Based on this fact, it has been concluded that the Bohm's dynamics is untenable as a physical theory \cite{S. Colin2014}. But it should be noted that  the D. Bohm did not consider the original version of his model as a final theory and  even suggested a general form for modifying the quantum force as follows \cite{D. Bohm52-1}:
\begin{equation}
m\frac{d^2{X}(t)}{dt^2}=-\nabla \left(V+Q\right)+{F}\left({p}-\nabla S\right),
\end{equation}
 however, he did not introduce any explicit form for the function ${F}\left({p}-\nabla S\right)$. In this paper, we derive a  natural generalized quantum force that leads to a stable  casual quantum theory. Moreover, we will show that in this modified Bohm's dynamics, a non-equilibrium distribution relaxes to equilibrium distribution as same as de Broglie's first-order dynamics. 

\emph{2. Modification of the Bohm's dynamics-}. Since the mathematical origin of the instability of the Bohm's equilibrium distribution is rooted in its singularity, we suggest replacing the Dirac delta function in eq.(5) by a Gaussian distribution with a width of $\mu> 0$ around $\nabla S$ :
\begin{equation}
f_{\mu }\left({x},{p};t\right)=R^2{\exp \frac{{({p}-\nabla S)}^2}{-\mu }\ }.
\end{equation}
Notice that, the above "modified equilibrium distribution", $f_\mu$, as same as the Bohm's  equilibrium distribution $ f_B $, is consistent with standard probabilistic interpretation of the wave function,  because:
\begin{equation}
\rho =\int{f_{\mu }\left({x},{p}\right)d^{3N}{p}}={\left|\psi \right|}^2,
\end{equation}
\begin{equation}
{J}=\int{({{p}}/m) f_{\mu }\left({x},{p}\right) d^{3N}{p}}={\left|\psi \right|}^2{\nabla S}/m.
\end{equation}
It is clear that, $f_\mu$  is not equivariant under Bohm's dynamics, so if we want to consider it as a quantum equilibrium distribution then the Bohm's dynamics must be modified. For this purpose, we derive the time evolution of the $f_\mu$, using Schr\"{o}dinger  equation:
\begin{equation}
\frac{\partial f_{\mu }}{\partial t}+\frac{{p}}{m}.\ \nabla f_{\mu }+{\nabla }_{{p}}.\left[\left(-\nabla V+{{F}}_Q\right)f_{\mu }\right]=0,
\end{equation}
Where ${\nabla }_{{p}}=({\nabla}_{{\bm p}_1} ,...,{\nabla}_{{\bm p}_N })$ is gradient operator in the  momentum space and  $F_Q $ is:
\begin{equation}
{{F}}_Q=-\nabla Q-\frac{\mu}{m} \frac{\nabla R}{R}+\frac{1}{m}\nabla \nabla (IS)\left(\nabla S-{p}\right),
\end{equation}
where $I$ is identity matrix in 3D position space. Equation (10) can be considered as a Liouville equation with an extra modifed quantum force $F_Q $.
So, we consider the particles' equation of motion as:
\begin{equation}
m\frac{d^2{{ X}}(t)}{dt^2}={(\left.{-\nabla V+{F}}_Q) \right|}_{\substack{x=X(t)\\p=P(t)}}, 
\end{equation}
Where $P=m{{\dot {X}}}$ is particles momentum vector. This "modified Bohm's dynamics" leads to the same Liouville equation as (10) for paticles distribution in phase space $ f(x,p;t)$. Therefor, if  $ f{\rm=}{f}_{\mu}$ at some initial time $t_0$ then $ f{\rm=}{f}_{\mu}$ for all future times (equivariance).
The equivariance of $ {f}_{\mu}$ in phase space ensures the equivariance of $ |\psi|^2$  in configuration space and  based on the causal theory of measurement [4], the consistency between statistical predictions of our model and standard quantum mechanics for all observables is guaranteed. 
  However, in our model the average  velocity of particles in each point is equal to ${\nabla} S/m$ , but the particles' trajectories are completely different from standard Bohemian trajectories, even if the initial velocities are chosen as $ {\dot {{X}}}(t_0){\rm=}{\nabla} S({X}(t_0),t_0)/m$. 
In fact, in equilibrium state, the standard Bohmian mechanics could be considered as $\mu \to0$ limit of our model: because in this limit, the modified quantum equilibrium distribution $ f_{\mu}$, reduces to the Bohm's quantum equilibrium distribution $f_B$, and so we have $P\to\nabla S$ for all particles in ensemble; Therefore in this case the modified quantum force reduces to standard Bohmian quantum force, ${F}_Q\to{\mathbf -}\nabla Q $.

As an example, consider the coherent states of a harmonic oscillator in one dimension, with$V(x)= k{x}^2/2 $:
\begin{equation}
\psi=N\exp{{\frac{-1}{2} ({(x-\alpha{\cos  t })}^2 +{i}(t+2xa{\sin  t}+{\alpha^2}{\sin  2t })})}
\end{equation}
where $N$ is normalization constant and $\alpha$ is an arbitrary constant. We choose $\hbar=k=m=1$, in this case, the eq.(12) leads to:
\begin{equation}
X\left(t\right)=\dot X_0\frac{{\sin  (\sqrt{\mu }t)\ }}{\sqrt{\mu }}+{\cos  (\sqrt{\mu }t)\ }(X_0-\alpha)+\alpha\ {\cos  t }
\end{equation}
Where $ X_0$ and $\dot X_0$ are initial position and velocity of particle respectively. In the limit $\mu\to 0$, the equation (14) reduces to corresponding equation of motion in Bohm's dynamics: $X\left(t\right)=X_0+\dot X_0t+\alpha\left({\cos  t}-1\right)$. In this case, as in [20] has been mentioned, if $ \dot X_0\neq1/m  (\partial S(x,0)/\partial x)$ Then the particle will escape to infinity and so the quantum equilibrium in Bohm's dynamics is unstable. But even for a small amount of $\mu$, our model circumvents this problem. In figure.1, particle trajectories in our model are compared with those of the Bohm's model. It is clear that this result in not limited to the above example and our model generally avoids the problem; because the modified equilibrium distribution $f_{\mu}$ includes the particles with $ {p} \neq \nabla S$.
\begin{figure}[b]
\includegraphics[width=0.45\textwidth, height=0.30\textwidth]{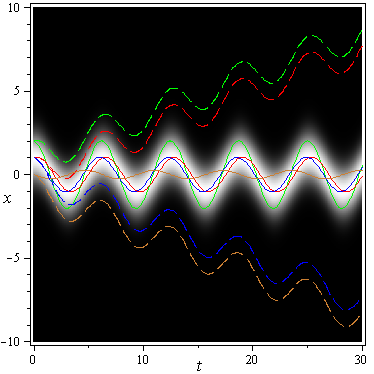}
\caption{\label{fig:epsart}The particle trajectories in our model (with ${\mu}={1}$) is compared by Bohm's trajectories in the case of the coherent state of harmonic oscillator (13), with $\alpha = 1$. Wave packet is represented by the White shaded area. The Bohm's trajectories are represented by dashed curves and particles trajectories in our model are represented by solid curves. All trajectories have initial velocities as $\dot X_0=\pm0.25\neq 1/m  (\partial S(x,0)/\partial x)$ and the trajectories with same color have same initial conditions.  The Bohm's trajectories escape the wave packet but the trajectories of our model follow the wave packet.}
\end{figure}

\emph{3. Sub-quantum H-theorem-.} In this section, we want to show that the non-equilibrium distributions approach to modified equilibrium distribution via an coarse-graining H-theorem similar Valentaini’s H-theorem. For this purpose, since our model is a second order dynamics, we must generalize Valentaini's H-function, 
\begin{equation}
H=\int{d^{3N} x  \ \rho  \ { \ln \left(\rho/|\psi|^2 \right)\ }},
\end{equation}
from configuration space  to phase space. For this, we introduce the counterpart quantities in the phase space as:
\begin{equation}
\rho(x) \to f(x,p) ,
\end{equation}
\begin{equation}
 |\psi(x)|^2  \to f_{\mu}(x,p) . 
\end{equation}
 So we redefine Valentini’s H-function in our model as :
\begin{equation}
H_{\mu }=\int{d\Omega \ f \ {\ln \xi  }},
\end{equation}
Where $d\Omega=d^{3N}x\ d^{3N}p$ is differential volume element in the phase space and $\xi$  is defined as $\xi{\rm=}f/f_{\mu}$. This H-function is a non-negative quantity and will be zero if and only if $f{\rm=}f_\mu$ everywhere, these features make it a useful measure of proximity to quantum equilibrium distribution. On the other hand, the $ \xi $ is preserved along particle’s trajectories in the phase space:
\begin{equation}
\frac{d}{dt}\xi (X(t),P(t);t)=0,
\end{equation} 
becuse $f$ and $f_{\mu}$ obey identical Liouville equations in our model. In addition, $fd\Omega$ is the number of the systems in ensemble which occupy a comoving volume $d\Omega$  and so it is preserved along trajectories too. Therefore the exact $H_\mu$ is constant in time, $dH_\mu/dt{\rm=}0$, and so a non-equilibrium distribution can never really relaxes to equilibrium distribution. But, since all physical measurements have a finite accuracy, it is reasonable to understand the relaxation process in terms of a coarse-grained H-function which isn't a conserved quantity, as same as classical  case \cite{Tolman1979,Davies1974}. For this purpose, we divide phase space into non overlapping cells of volume $\delta\Omega$ and define coarse-grained quantities as:
\begin{equation}
\bar f{\rm=}\frac{1}{\delta\Omega}{\int}_{\delta\Omega}{f d\Omega},
\end{equation}
and 
\begin{equation}
 {\bar f}_{\mu}{\rm=}\frac{1}{\delta\Omega}{\int}_{\delta\Omega}{{f}_{\mu} d\Omega},
\end{equation}
 where $\bar f$ and ${\bar f}_{\mu} $ are considered as constants in each cell. These values can be used to definitoin a course grained $H_μ$-function: 
\begin{equation}
{\overline{H}}_{\mu }=\int{d\Omega \ \overline{f}\ {\ln \overline{\xi } }},
\end{equation}
Where $\bar{\xi}$  is defined as $ \bar{\xi}{\rm=}\bar f / {\bar f}_{\mu}  $.
Now, to prove that $ {\bar H}_{\mu}$ decreases with time, same as the classical case, we assume no "microstructure" for the initial state, on the other word, we assume  ${\bar f}(t_0){\rm=}f(t_0)$ and  ${\bar f}_{\mu}̅(t_0){\rm=}{ f}_{\mu} (t_0)$. The proof now precedes along the lines of the Valentaini's proof for first order Bohemian mechanics:
From above assumption and the fact that the exact $ H_{\mu} $ is constant in time, we have;
\begin{equation}
{\overline{H}}_{\mu }\left(t_0\right)=H_{\mu }\left(t\right)=\int{d\Omega \ f_{\mu }\ \xi \ {\ln \xi  }}.
\end{equation}
In addition, by using the fact that $ \bar \xi $ is constant in each cell $\delta \Omega $, one can easily show that
\begin{equation}
{\overline{H}}_{\mu }\left(t\right)=\int{d\Omega \ \ f_{\mu }\xi \ {\ln \overline{\xi }\ }},
\end{equation}
and
\begin{equation}
\int{d\Omega \ \ f_{\mu }\left(\overline{\xi }-\xi \right)=0}.
\end{equation}
So by using equations (23),(24) and (25), we have:
\begin{equation}
{\overline{H}}_{\mu }\left(t_0\right)-{\overline{H}}_{\mu }(t)=\int{d\Omega \ \ f_{\mu }(\xi \ {\ln \left({\xi }/{\overline{\xi }}\right)\ }}+\overline{\xi }-\xi ).
\end{equation}
Finally, Since $ {x} \ { \ln(x/ y)}+y-x\ge 0$ for all $ x$, $y$, we have the subquantum H-theorem :
\begin{equation}
\frac{d{\overline{H}}_{\mu }}{dt}\le 0.
\end{equation}
The decrease of ${\bar H}_{\mu}$ shows relaxation of  non-equilibrium distribution  to modified equilibrium distribution $({\bar f}\to{\bar f}_{\mu})$. This coarse-grained H-theorem could be considered as the mathematical formulation of the Gibbs-Ehrenfest's idea in the context of the second-order Bohmian mechanics: a non-equilibrium phase space distribution will tend to increase fine-grained microstructure and become more like equilibrium distribution on a coarse-grained level. It can be considered as a statistical derivation of the Born rule and also "averaged guiding equation" in the context of the deterministic Bohm's approach, without postulating " random fluctuations", as done by Bohm and Vigier in \cite{D. Bohm1954}.

\emph{4. Discussion and concluding remarks-.} In this paper we introduce a modified quantum force which leads to a stable quantum equilibrium. But let us notice that the special form of the modified quantum equilibrium distribution $f_\mu$, that we used in this paper is not the only possible choice that could be used to make a stable second order dynamics and infact the only necessary properties for the modified quantum equilibrium distribution are as follows: (I) It must be a local function of $\psi$ that satisfies equations (8), (9) and (II) Lebesgue measure of its support sould be non-zero in phase space. As an example in definition of $f_\mu$, one can use the Lorentzian representation of the Dirac delta function instead of the Gaussian representation:
\begin{equation}
f_\mu (x,p)=\frac{R^2}{\pi} \frac{\mu}{(p-\nabla S)^2+{\mu}^2 },
\end{equation}
which leads to a different form for quantum force via Liouville equation. In fact, there are many  stable second order dynamics that lead to the same statistical predictions as standard quantum mechanics in the  equilibrium state.  But in principele for the non-equilibrium state, these modified dynamics may lead to different predictions that could be used as experimental test of them.  Since all ordinary  physical systems, such as atoms in the laboratory,  have had a long astrophysical history, we can not  see these systems in non-equilibrium state today. However, before ending  relaxation process in the early universe,  the non-equelibrium effects could be observed. In fact, it has been  proposed that the CMB data may be used to set bounds on non-equilibrium deviations from quantum theory \cite{A. Valentini2010, A. Valentini2007, S. Colin A. Valentini2013, S. Colin A. Valentini2015}. As an out look, the CMB data can be used to determine some limits on the value of the parameter $\mu$.
 \begin{acknowledgments}
The authors thank H. Abedi  and S. Y. Rokni for their useful comments and interesting discussions.
\end{acknowledgments}


\nocite{*}

\bibliography{apssamp}

\end{document}